\documentclass[twocolumn,pra,preprintnumbers,superscriptaddress,landscape,nofootinbib]{revtex4}
\usepackage[utf8]{inputenc}
\usepackage{amsmath}
\usepackage{amsthm}
\usepackage{amssymb}
\usepackage{float}
\usepackage{braket}
\usepackage{graphicx}
\usepackage{url}
\usepackage{here}
\usepackage{natbib}
\usepackage{rotating}
\usepackage{comment}
\usepackage[colorlinks=true, pdfstartview=FitV, linkcolor=red, citecolor=blue, urlcolor=blue]{hyperref}
\usepackage{slashed}
\usepackage{multirow}
\usepackage[normalem]{ulem}
\graphicspath{{./figures/}}

\newcommand{\be}{\begin{equation}}      
\newcommand{\ee}{\end{equation}}      
\newcommand{\bea}{\begin{eqnarray}}      
\newcommand{\eea}{\end{eqnarray}}

\newcommand{\Tr}{\mathrm{Tr}}

\newcommand{\ctext}[1]{\raise0.2ex\hbox{\textcircled{\scriptsize{#1}}}}
\usepackage{tikz}
\usetikzlibrary{quantikz}

\theoremstyle{definition}

\theoremstyle{remark}

\usepackage[normalem]{ulem}

\graphicspath{{./figures/}}
\begin{document}
\title{Exceeding the maximum classical energy density in fully charged quantum batteries}
\author{Masahiro Hotta}
\email[]{hotta@tuhep.phys.tohoku.ac.jp}
\affiliation{Department of Physics, Tohoku University, Sendai, Miyagi 980-8578, Japan}
\affiliation{Leung Center for Cosmology and Particle Astrophysics, National Taiwan University,
Taipei 10617, Taiwan, Republic of China}

\author{Kazuki Ikeda}
\email[]{kazuki.ikeda@umb.edu}
\affiliation{Department of Physics, University of Massachusetts Boston, Boston, MA 02125, USA}
\affiliation{Center for Nuclear Theory, Department of Physics and Astronomy, Stony Brook University, Stony Brook, New York 11794-3800, USA}
\affiliation{Co-design Center for Quantum Advantage, Department of Physics and Astronomy, Stony Brook University, Stony Brook, New York 11794-3800, USA}

\begin{abstract}
Quantum batteries are anticipated to achieve significant advancements in energy storage capacity. In classical batteries, the energy density at each subsystem reaches its maximum value, denoted as $E_C$, which is determined by dividing the maximum energy by the number of subsystems. We demonstrate that this limit can be surpassed in quantum batteries by protocols of Quantum Energy Teleportaion (QET), allowing for the energy density at a subsystem to exceed the value of $E_C$. Our protocol offers enhanced efficiency, reduces experimental complexity on quantum computers, and enables instantaneous energy charging through Local Operations and Classical Communication (LOCC). Leveraging quantum entanglement, this protocol significantly improves quantum energy storage systems, promising advances in quantum computing and new technological applications. This work represents a crucial step towards revolutionizing quantum energy storage and transfer.  
\end{abstract}

\maketitle
\section{
Quantum batteries }\cite{PhysRevE.87.042123,Binder_2015,PhysRevLett.118.150601,Campaioli:2023ndh,10.1116/5.0184903,PhysRevLett.128.140501,PhysRevA.108.042618,PhysRevLett.125.236402} represent a concept for energy storage devices utilizing many-body quantum systems, capable of achieving notably higher efficiency compared to conventional classical batteries. For example, multi-qubits systems are able to enhance the energy storage efficiency in power per qubit can be achieved when global operations are permitted \cite{Downing:2023nfa}. Such a quantum advantage may be applied to energy-saving technology in quantum computing, energy supply in quantum devices, and artificial quantum photosynthesis \cite{doi:10.1126/sciadv.abk3160}. \if{It is expected that quantum states have the potential for prolonged persistence of quantum batteries. For instance, time crystals have been demonstrated to persist for at least 40 minutes—approximately 10 million times longer than other known crystals \cite{NP2024}. }\fi Recently, it has been reported that a demonstration involving two Ca+ ions confined in a linear Paul trap successfully functioned as an energy conversion device, which can play a role of quantum batteries \cite{PhysRevLett132}. It is expected that the efficiency of quantum batteries is enhanced by quantum entanglement. Therefore, it is preferable to adopt a pure state with entanglement rather than a thermal mixed state where the entanglement is zero. This allows us to surpass the classical limit of energy density using a pure state where quantum entanglement remains even in the highest energy eigenstate.

Quantum entanglement is a great tool in quantum task and achieve high-ability in a various types of quantum tasks such as quantum computation and quantum communication. For instance, by using ground-state entanglement of many body systems, quantum energy teleportation (QET) is realized, which attains effective energy transportation only by local operations and classical communication (LOCC) \cite{HOTTA20085671,PhysRevD.78.045006,2009JPSJ...78c4001H}. Recently two independent QET experiments have been reported and gather much attention \cite{PhysRevLett.130.110801,2023arXiv230102666I}. This protocol harnesses the principles of quantum entanglement and feedback control, attracting significant interest across multiple disciplines. Its applications extend beyond quantum physics—including condensed matter and high-energy physics—into the realm of quantum communication~\cite{Ikeda:2023ljh,PhysRevD.107.L071502,https://doi.org/10.1049/qtc2.12090,10.1116/5.0164999,Ikeda:2024hbi,Ikeda:2023yhm,Wang:2024yqn,Itoh:2023pmj,10.1093/ptep/ptae192,ikeda2025quantum,Ikeda:2025gju}.

In this work, we aim to harness the potential of quantum entanglement and Local Operations with Classical Communication (LOCC) to establish an efficient protocol for quantum batteries. Specifically, we apply the QET protocol to enhance the performance of quantum batteries. Unlike classical batteries, where the energy density in each subsystem reaches a maximum value, ($E_C$), defined as the maximum energy divided by the number of subsystems, our results demonstrate that QET protocols can surpass this limit. This allows the energy density within a subsystem to exceed ($E_C$). 

In the conventional QET, the ground state of the system is used. Here we utilize the highest energy state $\ket{E_{\max}}$. Then our proposal can be summarized as follows:
\begin{itemize}
    \item[]``QET for $\ket{E_{\max}}$ is a quantum battery."
\end{itemize}

One of the significant advantages of our proposal is that the protocol has been experimentally verified and can be easily implemented using quantum hardware~(see the GitHub code \cite{Ikeda_Quantum_Energy_Teleportation_2023} for the implementation). While the technology shows tremendous promise, it may still be many years before it finds practical applications. Our research seeks to deepen the understanding and harness the principles of quantum mechanics to make these batteries practical in real-world scenarios.

In previous studies (see \cite{Campaioli:2023ndh}), the advantages of quantum entanglement in the context of quantum batteries have already been discussed. However, the role of local energy fluctuations induced by quantum entanglement has not been considered. In this paper, it is highlighted that quantum entanglement can play a crucial role in generating higher energy density within the subsystems of quantum batteries.

\section{Quantum Battery (review).}
Here we briefly review the recent advances and some features of quantum battery. Quantum batteries are defined as a quantum system capable of storing and releasing energy more efficiently than traditional batteries. Quantum batteries are theoretical devices that utilize quantum states, particularly quantum entanglement, to enhance energy storage and retrieval. They offer the potential for ultra-fast charging, nearly instantaneous under ideal conditions.

There are several theoretical progress in creating prototype quantum batteries, but they remain in the experimental stages. The key question addressed is how to maximize the work extractible from these batteries and the advantage of using entangling operations in achieving higher efficiency. Alicki and Fannes initiated the concept of quantum battery based on the passibity of quantum states~\cite{PhysRevE.87.042123}. A comprehensive overview of the recent study on quantum batteries is given in~\cite{Campaioli:2023ndh}. 

Recent studies have demonstrated the feasibility of these batteries on a small scale~\cite{PhysRevLett.131.240401}. The primary advantage is rapid charging, which could far exceed the capabilities of conventional batteries. Significant technical challenges need to be addressed, including maintaining the necessary quantum states and scaling the technology for practical use. Quantum coherence, the state in which quantum particles are entangled, is fragile and difficult to maintain over long periods and larger scales.

\section{Protocol of Quantum Battery based on QET} 
Here we describe our protocol in a spin chain consisting of locally interacting spin 1/2 particles (qubits), whose Hamiltonian can be decomposed into a sum of local Hamiltonians and its interaction term $V$ as follows:
\begin{equation}
    H=\sum_{n=0}^{N-1} H_n +V.
\end{equation}
The essential part of our protocol is called quantum energy teleportation~\cite{HOTTA20085671}. If we perform the following protocol to the sign-flipped Hamiltonian $-H$, then it corresponds to quantum energy teleportation, where the highest energy sate $\ket{E_{\max}}$ of $H$ is the ground state $\ket{E_{\min}}$ of $-H$. We assume that there is no degeneracy in the highest energy state, therefore $\ket{E_{\max}}$ is determined uniquely. Moreover we also assume that $\ket{E_{\max}}$ is entangled.

Our system consists of Alice and Bob, who share the highest energy state $\ket{E_{\max}}$ of the Hamiltonian $H$. First Alice performs a measurement of her Pauli operator $\sigma_{n_A}$ using her projective measurements operator $P_{n_A}(\mu)=(1+\mu \, \sigma_{n_A})/2$, resulting in either $\mu=-1$ or $\mu=+1$. When measuring the quantum state at subsystem $A$, Alice extracts energy from the system. (This is in contrast to QET, where Alice injects energy into the system by the measurement.) However, her operation does not affect Bob's local energy since the system has only local interactions and the condition $[P_A(\mu),H_{n_B}]=0$ is manifested. Through a classical channel, Alice transmits the result of her measurement $\mu\in\{-1,+1\}$ to Bob, who then implements an operation $U_{n_B}(\mu)$ on his qubit and separately measures his local Pauli operators $X_{n_B}$, $Y_{n_B}$, and $Z_{n_B}$. Following Bob's application of $U_{n_B}(\mu)$ to $P_{n_A}(\mu)\ket{g}$, the resulting density matrix is denoted as $\rho$:
\begin{equation}
    \rho=\sum_{\mu\in\{-1,1\}}U_{n_B}(\mu)P_{n_A}(\mu)\ket{E_{\max}}\bra{E_{\max}}P_{n_A}(\mu)U^\dagger_{n_B}(\mu). 
    \label{eq:rho_QET}
\end{equation}
Bob's localized system is calculated as $\langle E_{n_B}\rangle=\Tr[\rho H_{n_B}]$, which is not smaller than the Bob's initial local energy:
\begin{equation}
    \Tr[\rho H_{n_B}]\ge\Tr[\ket{E_{\max}}\bra{E_{\max}}H_{n_B}].
\end{equation}
Therefore Bob can charge more energy than he can originally store. In this sense, Bob can exceed the maximal energy storage and we call it quantum battery. This is in contrast to QET, where Bob's local energy can be even smaller than his local lowest energy because of the quantum feedback protocol.

Our mechanism for surpassing the classical upper bound of maximum energy density is independent of the specific design of the quantum battery. The presence of interaction terms plays a crucial role in enhancing energy fluctuations, in a manner analogous to how QET can generate regions of negative energy density. As illustrated in Figure 1, the charging energy is simultaneously injected into the battery, which consists of $N$ quantum subsystems. We assume here that the energy stored in each subsystem can be independently extracted. In the left panel of Figure 2, a classical scenario of a maximally charged battery is depicted. In contrast, the right panel of Figure 2 shows that the second subsystem carries an energy density exceeding the classical upper bound. This surplus energy can be directly extracted from the subsystem for practical use.

\begin{figure}
    \centering
    \includegraphics[width=\linewidth]{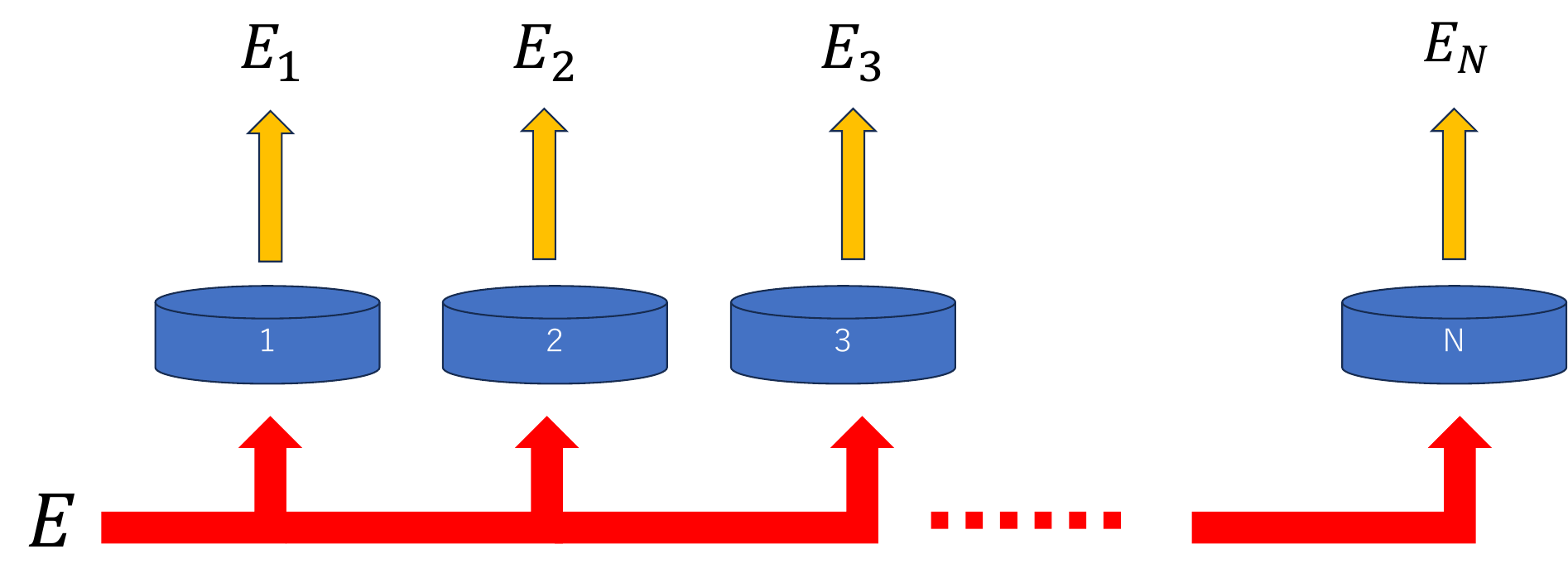}
    \caption{The concept of a quantum battery in our scheme is illustrated. The battery consists of $N$ subsystems, and energy $E$ is stored simultaneously across all of them. Energy can be extracted individually from each subsystem. In particular, For instance, if subsystem 2 holds energy exceeding the classical bound, it is possible to extract that energy, $E_2$ specifically from subsystem 2.}
    \label{fig:Fig1}
\end{figure}

\begin{figure}
    \centering
    \includegraphics[width=\linewidth]{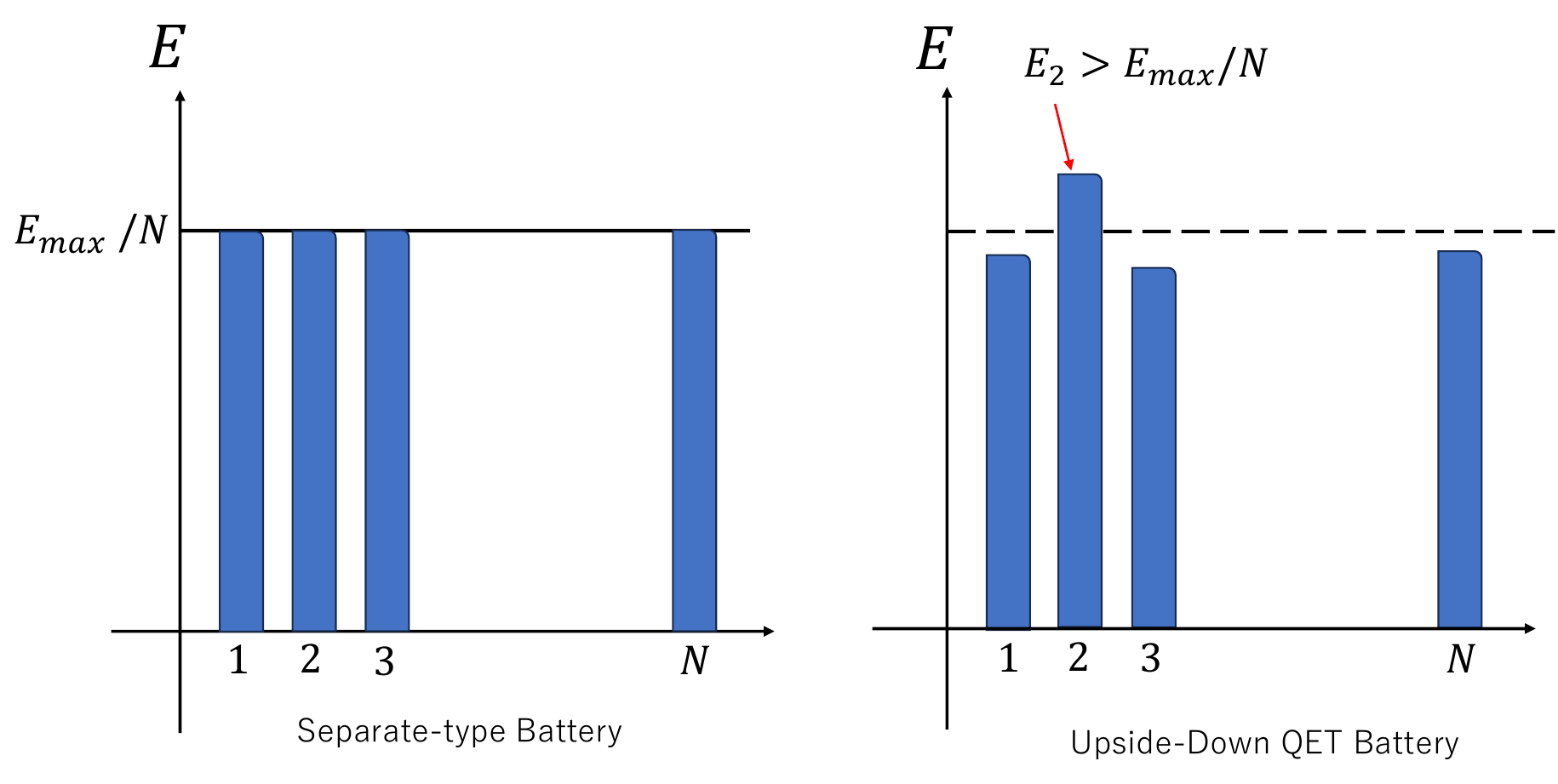}
    \caption{A case of energy storage by the battery is illustrated. The amount of energy $E_2$ stored in subsystem 2 exceeds the classical bound $E_{max}/N$.}
    \label{fig:Fig1}
\end{figure}

In what follows we describe the Bob's optimized operation that allows him to store the maximal energy to his local system. For the convenience of discussion, we normalize Bob's maximal energy to 0: $\Tr[\ket{E_{\max}}\bra{E_{\max}}H_{n_B}]=0$. This can be always done by adding a local constant to his energy.  

First, Bob's unitary operation $U_{n_B}(\mu)$ is given by
\begin{equation}
    U_{n_B}(\mu)=\cos\theta I-i\mu\sin\theta\sigma_{n_B},
\end{equation}
where $\theta$ satisfies
\begin{align}
    \cos(2\theta)=\frac{\xi}{\sqrt{\xi^2+\eta^2}}\,, \quad
    \sin(2\theta)=\frac{\eta}{\sqrt{\xi^2+\eta^2}}\,,
\end{align}
with
\begin{align}
\begin{aligned}
\xi&=\bra{E_{\max}}\sigma_{n_B}H\sigma_{n_B}\ket{E_{\max}}\\ \eta&=\bra{E_{\max}}\sigma_{n_A}\dot{\sigma}_{n_B}\ket{E_{\max}}\\
\dot{\sigma}_{n_B}&=i[H,\sigma_{n_B}]\,.
\end{aligned}
\end{align}
The local Hamiltonian must be selected in such a way that $[H,\sigma_{n_B}]=[H_{n_B},\sigma_{n_B}]$. After Bob applies the operation $U_{n_{B}}(\mu)$ to $P_{n_A}(\mu)\ket{g}$, the mixed state $\rho$ is obtained in average. Subsequently, the average energy that Bob can charge is
\begin{equation}
\label{eq:QET}
    \langle E_{n_B}\rangle=\Tr\left[\rho H_{n_B}\right]=\frac{1}{2}\left[\sqrt{\xi^2+\eta^2}-\xi\right]\ge0\,.
\end{equation}
Here it is important that $\langle E_{n_B}\rangle$ is positive when $\eta$ is no zero. Remember that $\langle E_{n_B}\rangle$ is non-positive for the QET, which distinguishes our protocol of quantum battery from QET. 

Before proceeding to the next discussion, let us briefly explain how our proposal differs from conventional studies~\cite{PhysRevE.87.042123,Binder_2015,PhysRevLett.118.150601,Campaioli:2023ndh,10.1116/5.0184903,PhysRevLett.128.140501,PhysRevA.108.042618,PhysRevLett.125.236402}. First of all, the experimental verification of the QET protocol is done by quantum hardware~\cite{PhysRevLett.130.110801,2023arXiv230102666I,Ikeda:2024hbi,Ikeda_Quantum_Energy_Teleportation_2023} and applying this to our quantum battery is straightforward. It is crucial to remember that the experimental realization of a quantum battery remains one of the most challenging tasks. However, our protocol effectively addresses and resolves this issue, thereby making the technology more accessible and less cumbersome compared to traditional methods. Moreover leveraging LOCC, our protocol allows for instantaneous energy charging, a feature not present in current methodologies.

\section{Model and Results}
Here we explain our model that we used for the demonstration, inspired from the minimal model of QET~\cite{2011arXiv1101.3954H,2010PhLA..374.3416H}. Our protocol works for a generic locally interacting model, however, to work on a concrete setup, we use the minimal model. Our results can be easily testable using a real quantum device, by simply flipping the sign of the Hamiltonian~\eqref{eq:Ham} ($H\to -H$), by which the protocol corresponds to the minimal QET. See ~\cite{2023arXiv230102666I,Ikeda:2024hbi,ikeda2025quantum} for the recent work using IBM Quantum Computers, which offers a free access to up to 127 qubits for the generic public.   

The Hamiltonian of the minimal model is given by 
\begin{align}
\begin{aligned}
\label{eq:Ham}
    H&=H_0+H_1+V,\\
    H_n&=-hZ_n-\frac{h^2}{\sqrt{h^2+k^2}},~(n=0,1)\\
    V&=-2kX_0X_1-\frac{2k^2}{\sqrt{h^2+k^2}},
\end{aligned}
\end{align}
where $k,h$ are real numbers. In what follows we assume that $k$ is not 0 so that the system maintains entanglement in the ground state. Then we define Alice and Bob's local Hamiltonians as $H_A=H_0$ and $H_B=H_1+V$, respectively. The constant terms in eq.~\eqref{eq:Ham} are added so that it makes easier to understand the non-triviality of the protocol. In fact, the maximal energy state $\ket{E_{\max}}$ of $H$ can be obtained analytically: 
\begin{equation}
\label{eq:groundstate}
    \ket{E_{\max}}=\frac{1}{\sqrt{2}}\sqrt{1-\frac{h}{\sqrt{h^2+k^2}}}\ket{00}-\frac{1}{\sqrt{2}}\sqrt{1+\frac{h}{\sqrt{h^2+k^2}}}\ket{11}.
\end{equation}
It is easy to check analytically that the following conditions are maintained: 
\begin{align}
\begin{aligned}
\label{eq:normalize}
\bra{E_{\max}}H\ket{E_{\max}}&=\bra{E_{\max}}H_n\ket{E_{\max}}\\
&=\bra{E_{\max}}V\ket{E_{\max}}=0.
\end{aligned}
\end{align}
These equations indicate that it is impossible to increase the energy to a value greater than zero through any unitary operation. Nevertheless, Bob is capable of surpassing his local maximum energy threshold as a result of Alice's measurement, which disrupts unitarity and leads to the formation of the mixed state \eqref{eq:rho_QET} through such a non-unitary procedure. 

This argument reflects the core concepts of QET, with one significant difference highlighted below. In the context of QET, Bob can harness energy without actively engaging in the protocol, simply by allowing sufficient time for natural system evolution. Conversely, in this scenario of the quantum battery, he is unable to accumulate additional energy because $\bra{E_{\max}}U^\dagger H_{B}U\ket{E_{\max}}$ cannot be greater than 0 for any unitary $U$ regardless of the duration he waits.

Although Alice and Bob directly interact with each other through the $X_0X_1$ term, we can still ensure the non-triviality of the protocol by having Alice use $\sigma_A=X_0$ to extract energy from her system. Here it is extremely important that $X_A$ commutes with  $H_B$, ensuring that Alice's measurement dose not affect Bob's local energy at all.

To the highest energy state $\ket{E_{\max}}$, Alice performs her projective measurement in the $X$-basis and, as soon as she observes $\mu\in\{-1,1\}$, she tells her result to Bob, who operates $U_B(\mu)$ to his qubit and measures his energy. Here $U_B(\mu)$ is 
\begin{equation}
    U_B(\mu)=\cos\phi I-i\mu\sin\phi Y_B,
\end{equation}
where $\phi$ can be any real parameter. In our case we take it as 
\begin{align}
\begin{aligned}
    \cos(2\phi)&=\frac{h^2+2k^2}{\sqrt{(h^2+2k^2)^2+h^2k^2}}\\
    \sin(2\phi)&=\frac{hk}{\sqrt{(h^2+2k^2)^2+h^2k^2}}.
\end{aligned}
\end{align}
This choice of $\phi$ optimize the energy. When Alice and Bob repeat this procedure a sufficient number of times, the average quantum state $\rho$ eq.\eqref{eq:rho_QET} is obtained. The average energy that Bob can charge is obtained as
\begin{align}
\begin{aligned}
\label{eq:ana}
    \langle E_B\rangle&=\sum_{\mu\in\{-1,1\}}\bra{E_{\max}}P_A(\mu)U^\dagger_B(\mu)H_BU_B(\mu)P_A(\mu)\ket{E_{\max}}\\
    &=\frac{1}{\sqrt{h^2+k^2}}[hk\sin(2\phi)-(h^2+2k^2)(1-\cos(2\phi))]\\
    &>0.
\end{aligned}
\end{align}
As opposed to eq.~\eqref{eq:normalize}, Bob's local energy after the feedback from Alice is now positive. In this sense, Bob can exceed the maximal energy storage, which cannot be achieved with the quantum feedback-control. This is the exactly same as the energy that Bob can gain from the system by QET with respect to the ground state of $-H$. See eq.~(14) in \cite{2011arXiv1101.3954H} and the Appendix of \cite{2023arXiv230102666I} for the numerics.

\section{Conclusion}
In this work, we initiated a new protocol for a Quantum Battery, inspired by the concept of Quantum Energy Teleportation. For the purpose of simplicity, we limited ourselves to the minimal model. Our discussion can be easily extended to include a wide variety of locally interacting spin chain systems. Compared with the existing methods, the advantages of our protocol are three-fold: it not only enhances efficiency to implement a quantum battery but also significantly reduces the complexity of the experimental implementation process (at least on quantum computers), making it more practical and effective for real-world applications. Moreover, our protocol allows one to charge the energy instantly because it relies on Local Operations and Classical Communication (LOCC).

By leveraging quantum entanglement, we can achieve unprecedented efficiency and instantaneous energy transfer, thus paving the way for more advanced and practical quantum energy storage systems. This advancement not only promises substantial improvements in the field of quantum computing but also opens up new possibilities for future technological applications. Further research should be directed towards exploring the scalability of our protocol in larger and more complex quantum systems. Additionally, experimental validation using real-world quantum devices will be crucial to demonstrate the practicality and reliability of our theoretical results, although the essential part of the experiment has been carried out already.

In conclusion, our novel protocol represents a significant step forward in the development of quantum batteries. The principles and techniques introduced herein hold the potential to revolutionize energy storage and transfer in quantum technologies. We hope that this work will inspire continued exploration and innovation within the field, ultimately contributing to the advancement and application of quantum science and technology.

\section*{Acknowledgements}
MH is supported by Grant-in-Aid for Scientific Research (Grant No.21H05188, 21H05182, and
JP19K03838) from the Ministry of Education, Culture, Sports, Science, and Technology (MEXT), Japan, by an OIST SHINKA project, and by the Leung Center for Cosmology and Particle Astrophysics (LeCosPA), National Taiwan University. KI is supported by the U.S. Department of Energy, Office of Science, National Quantum Information Science Research Centers, Co-design Center for Quantum Advantage (C2QA) under contract number DE-SC0012704.

\section*{References}
\bibliographystyle{utphys}
\bibliography{main}
\end{document}